\documentclass[twocolumn,preprintnumbers,nofootinbib,prd,superscriptaddress,aps]{revtex4-1}

\usepackage[utf8]{inputenc} 
\usepackage{graphicx,amssymb,amsmath,amsthm,amsfonts,epstopdf,epsfig,epsf,times}
\usepackage[linktocpage]{hyperref}
\usepackage[usenames]{color}
\usepackage{epstopdf}

\graphicspath{{plots/}} 

\usepackage{textcomp}
\usepackage{bm}
\usepackage{dcolumn}
\usepackage{latexsym}
\usepackage{rotating}
\usepackage{hyperref}
\usepackage{color}
\usepackage{longtable}
\usepackage{enumerate}
\usepackage{tensor}
\usepackage{stmaryrd}
\usepackage[normalem]{ulem}

\usepackage{mathtools}
\usepackage{url}
\begin{document}\title {\large Study of $t\bar{t}H$ production with $H\rightarrow b\bar{b}$ at the HL-LHC}

\author{A. J. Costa}
\affiliation{School of Physics and Astronomy, University of Birmingham, Edgbaston Park Rd, Birmingham B15 2TT, United Kingdom}
\author{A. L. Carvalho}
\affiliation{LIP, Av. Prof. Gama Pinto, 2, 1649-003 Lisboa, Portugal}
\author{R. Gon\c{c}alo}
\affiliation{LIP, Av. Prof. Gama Pinto, 2, 1649-003 Lisboa, Portugal}
\affiliation{Faculdade de Ci\^{e}ncias da Universidade de Lisboa, Campo Grande 016, 1749-016 Lisboa, Portugal}
\author{P. Mui\~{n}o}
\affiliation{LIP, Av. Prof. Gama Pinto, 2, 1649-003 Lisboa, Portugal}
\affiliation{Departamento de F\'{\i}sica, Instituto Superior T\'ecnico -- IST, Universidade de Lisboa -- UL, Avenida Rovisco Pais 1, 1049 Lisboa, Portugal}
\author{A. Onofre}
\affiliation{LIP, Departamento de Física, Universidade do Minho, 4710-057 Braga, Portugal}

\begin{abstract}
A feasibility study for an experimental analysis searching for $t\bar{t}H(H\rightarrow b\bar{b})$ production at the LHC and its high luminosity phase is presented in this note. Unlike search strategies currently being used in experimental collaborations, the present analysis exploits jet substructure techniques and focuses on the reconstruction of boosted Higgs bosons, to obtain sensitivity to the signal in a simple cut-based analysis. The $t\bar{t} +$ jets background may be constrained in the proposed analysis through a control region with very small signal contamination. Using this analysis strategy, the $t\bar{t}H(H\rightarrow b\bar{b})$ process could be observed at the LHC, in the semi-leptonic channel alone, with a significance of $5.41\pm 0.12$ for $\mathcal{L}=300\,\mbox{fb}^{-1}$. For the same integrated luminosity, in the High Luminosity LHC scenario with an upgraded detector, a significance of $6.13\pm 0.11$ may be obtained. The top Yukawa coupling could be measured with  a 35\% uncertainty using $\mathcal{L}=300\,\mbox{fb}^{-1}$ of LHC data and of 17\% at the HL-LHC scenario with $\mathcal{L}=3000\,\mbox{fb}^{-1}$. In the same luminosity scenarios, the signal strength is equally expected to have a 18$\%$ and 5$\%$ uncertainty, respectively. Finally, it was found that re-clustered jets may be used without loss of efficiency.  
\end{abstract}

\date{\today}

\maketitle


\section{Introduction}\label{sec:intro}

An impressive amount of work has been devoted to measuring the Higgs boson properties since its discovery in 2012, by the ATLAS and CMS experiments~\cite{higgsAtlas,higgsCMS} at the LHC. The interaction of the Higgs boson with fermions is one of its most important features, as it is responsible for their masses. Of these, the most experimentally accessible Yukawa couplings are to third generation quarks and leptons.

Both ATLAS and CMS have recently observed the coupling of the Higgs boson to top~\cite{ttHobs,ttHobs2} and bottom~\cite{bbobs,bbobs2} quarks, and to tau leptons~\cite{tauobs,tauobs2}. Of these, the top quark Yukawa coupling is of particular relevance, due to the large top quark mass, and might reveal a window into the physics beyond the Standard Model (SM). 

The production of the Higgs boson in association with two top quarks, $t\bar{t}H$, is especially important, as it provides direct experimental access to the $t\bar{t}H$ vertex at Born level. 
However, this process contributes only around $1\%$ of the total Higgs boson production cross-section, due to the large invariant mass of the final state objects. Nevertheless, different Higgs decay modes are accessible in this process, and the decay of the Higgs boson into two bottom ($b$) quarks poses an interesting scenario, as this decay is associated to the largest branching ratio of the Higgs particle ($58\%$)~\cite{brcs}, and contributes to a distinctive experimental signature. 

In the present work, we have considered the final state where one top decays hadronically and the other semi-leptonically, and the Higgs boson decays to a $b$-quark pair ($t\bar{t}H(H\rightarrow b\bar{b})$). Despite the distinct final state, this channel is dominated by large systematic uncertainties, related to a poor understanding of the production of $t\bar{t}$ pairs in association with heavy-flavour quarks (bottom or charm). This leads to a reasonably poor experimental sensitivity (see e.g.~\cite{ttH}).

The present paper proposes an alternative strategy for experimental $t\bar{t}H(H\rightarrow b\bar{b})$ searches at the High-Luminosity LHC (HL-LHC), based on a strategy proposed for the Future Circular Collider~\cite{mangano}. It relies on the reconstruction of Higgs bosons with high transverse momentum ($p_T$), so that its decay products are confined in a large radius jet. Hadronic jet substructure information is used to further discriminate between signal and backgrounds. 





The high luminosity phase of the LHC~\cite{HL,HL3} is expected to start operation in 2026 and run for ten years, collecting up to $3 - 4\ \rm{ab^{-1}}$ of proton collisions at a center-of-mass energy of 14 TeV. 
The number of expected collisions per bunch crossing, or {\it pileup}, will be up to 200, much higher than at present. 

\section{Simulation}
\label{sec:sim}

In addition to the signal and its main irreducible background, this paper also considered other relevant background processes, namely $t\bar{t}Z$, $t\bar{t}j$, where $j$ corresponds to additional jets,
$W^{\pm}b\bar{b}$, $b\bar{b}j$
and QCD di-jet production.
An alternative $t\bar{t}A$ signal sample was also generated with the \textsc{HC\_UFO\_v4.1} model~\cite{ttamodel}, where $A$ is a pure pseudo-scalar boson instead of the SM scalar Higgs. 

Events for these processes were generated at Leading Order (LO), and for a center-of-mass energy of 14 TeV. The \textsc{MG5\_aMC@NLO} generator \cite{mg5} and the LO \textsc{NN23LO1} PDF were used for all samples except for the di-jet sample, which was generated with \textsc{Pythia8.2}~\cite{pythia} using the LO \textsc{CTEQ 5L} PDF. \textsc{MadSpin}~\cite{madspin} was used in the generation of all \textsc{MG5\_aMC@NLO} samples, to preserve spin information in particle decays.


At generator level, cuts were applied to enhance the generation efficiency, and their effect on the analysis outcome was verified to be negligible. Leptons and $b$ quarks were required to have a minimum transverse momentum of $10\ \rm{GeV}$ in all samples with the exception of the di-jet sample (where a $b$-quark transverse momemtum cut was applied ($p_{T,b}>300\ \rm{GeV}$), and the $b\bar{b}j$ sample ($p_{T,b}>20$ GeV). 

Non-$b$-initiated jets were required to have transverse momenta $p_{T,j}>10$ GeV in the $t\bar{t}H$, $t\bar{t}b\bar{b}$ and $t\bar{t}Z$ processes. 
On the other hand, jets were required to satisfy $p_{T,j}>100\ \rm{GeV}$ in the $t\bar{t}j$ sample and  $p_{T,j}>50\ \rm{GeV}$ in the generation of the $b\bar{b}j$ sample. The minimum $p_{T,j}$ cut was set to $300\ \rm{GeV}$ for the di-jet sample. Furthermore, a minimum angular separation between pairs of jets and leptons $\Delta R_{jj,bb,jl} > 0.1$ was required, with $\Delta R_{ik}=\sqrt{(\eta_i-\eta_k)^2+(\phi_i-\phi_k)^2}$.

All simulated events are hadronized using \textsc{Pythia8.2}, and \textsc{Delphes3.2}~\cite{delphes} is used for the fast simulation of the collider experiments. The ATLAS default card was considered for simulating the LHC scenario, while the HL-LHC card was used for higher luminosity scenarios. 

Finally, in \textsc{Delphes}, simulated leptons are required to have a minimum transverse momentum $p_T$ of $10\ \rm{GeV}$, and an isolation variable $I$ below 0.1 within $\Delta R<0.3$, meaning that the $p_T$ of a $R = 0.3$ jet around the lepton must be less than $10\%$ of the lepton $p_T$, in order to consider it an isolated lepton.

\section{$b$ Tagging}
\label{sec:btagging}

The identification of $b$-quark initiated jets, or $b$-tagging, was emulated by searching for a $b$ quark within $\Delta R = 0.3$ of each jet. In the LHC (HL-LHC) scenarios, a $b$ quark was found, the jet was considered $b$-tagged with a probability of 61\% (65\%). Otherwise, a $c$ quark was sought for and, if found, a 4.5\% (3\%) probability was assigned for mis-tagging this jet as a $b$-jet. Finally, a 0.08\% (0.07\%) was assigned to mis-tagging jets initiated by light quarks or gluons. These working points were determined from existing literature for the HL-LHC~\cite{pixelbtag} and LHC scenarios, and are summarised in Table~\ref{wpbtag}.

$b$-tagged jets are required to have $\eta<2.5$. Improvements in the $b$-tagging in the HL-LHC scenario are only expected within this range, with performance improvements beyond this range still under optimization and uncertain.

\begin{table}[h!]
\centering
\caption{$b$-tagging working points. $q$ stands for quark and 'prob' for probability. $\varepsilon$ is the efficiency. $light$ quarks are $u$, $d$ and $s$ quarks.}
\vspace{5pt}
\begin{tabular}{c|c|c|c}
\hline
\hline
Scenario & \small{$b$-tag $\varepsilon$} & \small{$c$-tag $\varepsilon$} & \footnotesize{$ light\,q$ mistag prob} \\
\hline
\hline
LHC & $61\%$ & $4.5\%$ & $0.08\%$ \\
HL-LHC & $65\%$ & $3\%$ & $0.07\%$ \\
\hline
\hline
\end{tabular}
\label{wpbtag}
\end{table}

\section{Event Selection}

The analysis proposed in the present study corresponds to adapting and optimizing for the HL-LHC the strategy proposed in Ref.~\cite{mangano}. The main differences are explained below.

Selected events are required to have an isolated charged lepton ($e$ or $\mu$), with $p_T>30\ \rm{GeV}$ and $|\eta|<2.5$. To avoid the need for unfeasibly large samples, the isolated charged lepton is not required for the $b\bar{b}j$ and di-jet backgrounds. Instead, one in $5\times10^3$ jets is identified as a lepton, to emulate fake lepton identification. 

The calorimeter towers in the simulated event are then collected and the ones within a $\Delta R<0.1$ of an isolated electron are removed to avoid double-counting energy deposits. Muon energy deposits in the calorimeter are considered negligible. The remaining towers form the 'tower collection' and are used as input to jet clustering, done with \textsc{Fastjet}. The Cambridge-Aachen (C/A)~\cite{ca} algorithm is used to reconstruct jets with a radius $R=1.2$ and $p_T>180\ \rm{GeV}$. One or more of these jets are required.

The $R=1.2$ C/A jets are used to search for Higgs boson candidates using the BDRS tagger~\cite{bdrs}. This algorithm attempts to identify jets containing two sub-jets and a significant invariant mass. The algorithm parameters were a mass drop condition of 0.9 and $y_{cut} = 0.09$. If the algorithm identifies a Higgs candidate jet, its two subjets are required to be $b$-tagged, and have $p_T>30\ \rm{GeV}$. 

The candidates that pass these selection criteria are then filtered~\cite{bdrs} to remove eventual pile-up and underlying event contamination. Up to three hard thinner subjets are kept, to account for gluon radiation of one of the $b$ quarks. After this procedure candidate jets are required to have $p_T>180\ \rm{GeV}$.

Higgs candidate jets with a $\Delta R$ between the two BDRS $b$-tagged sub-jets ($\Delta R_{bb}$) below 0.3 are rejected, in order to suppress wrongly identified Higgs candidates. As a side-effect, the low-$\Delta R$ events provide a useful side-band at low jet masses. 

In events with more than one Higgs jet candidate (around $1\%$ of events) the jet with highest $p_T$ is chosen. The event is then required to have one Higgs candidate, and its associated towers are removed from the tower collection to avoid energy double counting in subsequent steps.

The remaining towers are clustered in $R=0.4$ anti$-k_t$ jets, which are required to have $p_T>30\ \rm{GeV}$. Two $b$-tagged jets are then required, with $\Delta R>0.4$ between them.

The $\Delta R$ between the leading and sub-leading $b$-tagged jets, and the Higgs candidate jet are computed and referred to as $\Delta R_{b_3,H}$ and $\Delta R_{b_4,H}$, respectively. Events are then required to satisfy $0.36\le \Delta R_{bb}\le 1.28$, $\Delta R_{b_3,H}\ge 0.87$ and $\Delta R_{b_4,H}\ge 0.88$, to suppress backgrounds. 

The main changes with respect to the original analysis~\cite{mangano} are that no use is made of the \textsc{HEPTopTagger2}~\cite{toptag} algorithm to tag hadronically decaying top quarks, since this was found to suppress the signal efficiency in the kinematic regime of the HL-LHC, and also in the C/A jet radius and jet $p_T$ cuts. Comparing both strategies when applied to HL-LHC simulated events, the proposed analysis corresponds to a factor $\sim 3$ improvement in the analysis significance.

\begin{figure}[h!] 
\centering    
\includegraphics[scale=0.4]{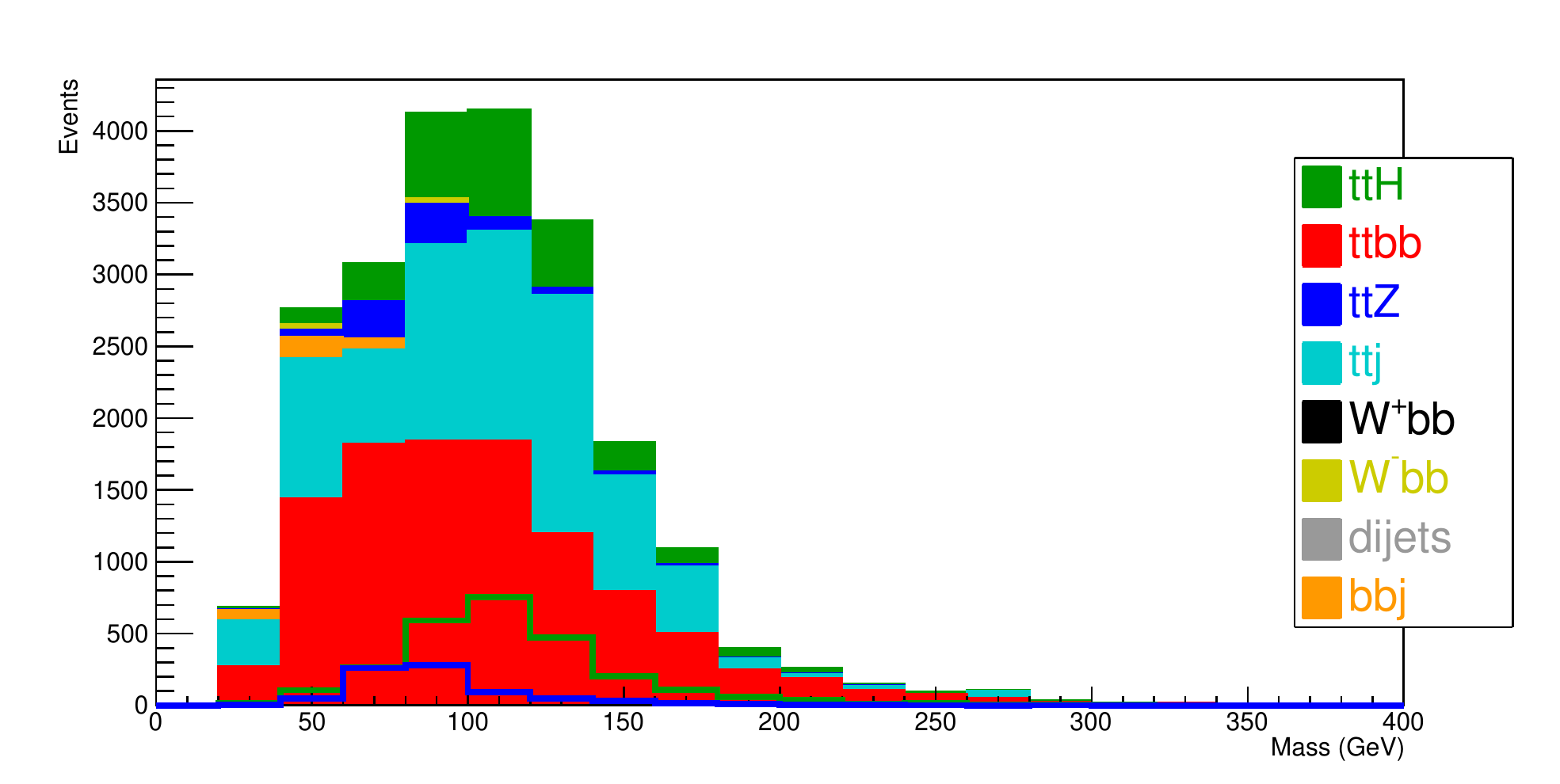}
\caption{Higgs candidates mass for $t\bar{t}H$ and backgrounds, for optimized analysis strategy. Events are normalized to $\mathcal{L} = 3000 \,\mbox{fb}^{-1}$.}
\label{mass_final}
\end{figure}

The significance and signal to background ratio was determined for Higgs candidate jets with mass, $m_H$, between 60 and 160 $\rm{GeV}$. Moreover, the significance is computed using $S/\sqrt{B}$. The mass distribution of the Higgs candidate jets, for the HL-LHC scenario and SM samples, is shown in Figure \ref{mass_final}, for integrated luminosity of 3000 $\mbox{fb}^{-1}$.

The estimated significance and $S/B$, estimated in the mass window between 60 and 160 GeV, are shown in Table~\ref{final_sig} for different integrated luminosities.

\begin{table}[h!]
\centering
\caption{Significance and S/B for different integrated luminosities, computed from Higgs candidate mass in range $[60,160]\ \rm{GeV}$.}
\vspace{5pt}
\begin{tabular}{c|c|c}
\hline
\hline
$\mathcal{L}$ ($\mbox{fb}^{-1}$) & $S/\sqrt{B}$ & $S/B$ ($\%$)\\ 
\hline
\hline
36  & 2.12 $\pm$ 0.04 & 15.7 $\pm$ 0.4 \\
300 & 6.13 $\pm$ 0.11 & 15.7 $\pm$ 0.4 \\
3000 & 19.39 $\pm$ 0.33 & 15.7 $\pm$ 0.4 \\
\hline
\hline
\end{tabular}
\label{final_sig}
\end{table}

\section{\textbf{Jet Re-Clustering}}

Jet re-clustering corresponds to using standard, $R=0.4$ anti-$k_t$ jets as input to jet clustering algorithms. In addition to good noise suppression characteristics, a further practical advantage of using re-clustered jets is to avoid maintaining many dedicated calibrations for each combination of jet algorithm parameters. 

To study the effect of jet re-clustering, we used $R=0.4$ anti$-k_t$ jets as input to the C/A jet reconstruction algorithm, with $R=1.2$, and applied a $p_T>180\ \rm{GeV}$ cut to the resulting jets, before using them as input to the BDRS Higgs tagger as in the analysis described above.

Apart from statistical fluctuations, no significant differences were found betwen the analyses with tower jets or re-clustered jets, neither in the shape of the invariant mass distribution nor in the significance.

\section{Control Region}

\begin{figure}[h!]
\centering
\includegraphics[scale=0.4]{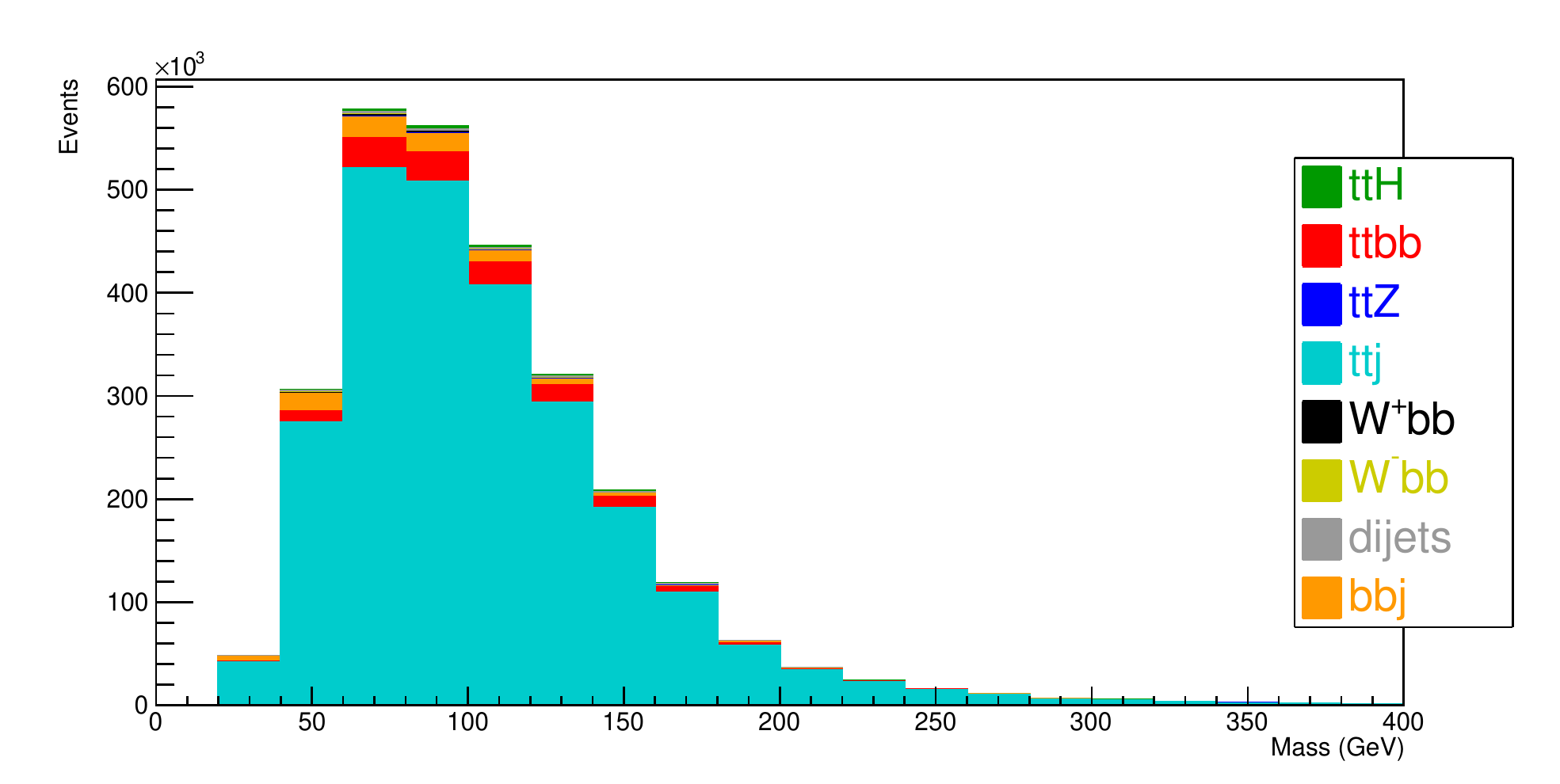}
\caption{Invariant mass distribution of Higgs candidates in the $t\bar{t}H$ control region. Events are normalized to $\mathcal{L} = 3000 \,\mbox{fb}^{-1}$.}
\label{mass_control}
\end{figure}

A control region is proposed, which may be used to constrain the $t\bar{t}j$ background normalization in the signal region. 
This control region is defined by an event selection which is identical to the signal region, except that the two Higgs candidate sub-jets are anti-$b$-tagged. The probabilities associated to requiring two $b$-tags on the two subjets of the Higgs candidate jet, retrieved by the BDRS Higgs tagger, are complementary of the working point used in the signal region. This results in the invariant mass distribution shown in Figure~\ref{mass_control} for the HL-LHC scenario. As expected, this region is dominated by the $t\bar{t}j$ background, with signal accounting for only 0.5\% of the event yield in the $[60,160]\ \rm{GeV}$ mass region.

\section{LHC Scenario}

\begin{figure}[h!] 
\centering    
\includegraphics[scale=0.4]{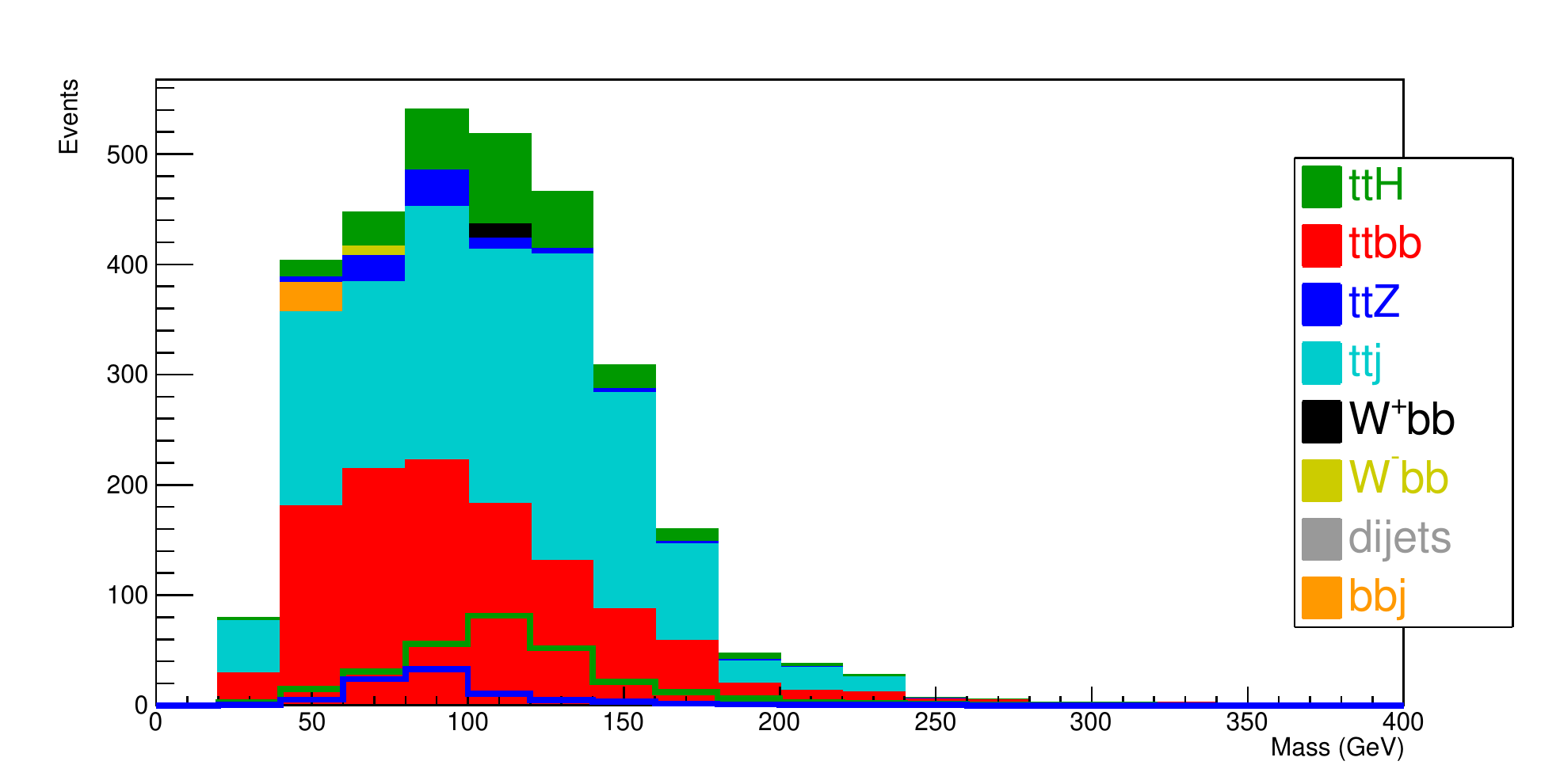}
\caption{Invariant mass distribution of Higgs candidates in the LHC scenario for $\mathcal{L} = 300 \,\mbox{fb}^{-1}$.}
\label{mass_lhc}
\end{figure}

The same analysis selection optimized for the HL-LHC scenario, i.e. using the HL-LHC simulation and assuming $3\ \mbox{ab}^{-1}$ of integrated luminosity, was then applied to the LHC scenario, where the ATLAS detector (fast simulation) and $36$ or $300\ \mbox{fb}^{-1}$ were assumed.
The ATLAS fast simulation model approximates the current ATLAS detector. Notable differences with respect to the HL-LHC simulation are
a slightly less performant $b$-tagging and a 2 T magnetic field (instead of a 3 T field in the HL-LHC case~\cite{delphes}). The mass distribution obtained for the LHC scenario is presented on Figure~\ref{mass_lhc} for an integrated luminosity of 300 $\mbox{fb}^{-1}$. 

The significance and $S/B$ of the analysis (optimized for the HL-LHC and applied to both scenarios) is shown for different integrated luminosities in Table \ref{lhc_sig}. As before, these variables are computed in the mass window between 60 and 160 GeV. 

\begin{table}[h!]
\centering
\caption{Significance and S/B ratio computed in the range $[60,160]\ \rm{GeV}$ for different integrated luminosities and detector simulations. The uncertainties are statistical and come from aassuming an uncertainty of $\sqrt{N}$ in the content of each bin.
}
\vspace{5pt}
\begin{tabular}{c|c|c|c}
\hline
\hline
Scenario & $\mathcal{L}$ ($\mbox{fb}^{-1}$) & $S/\sqrt{B}$ & $S/B$ ($\%$)\\ 
\hline
\hline
LHC & 36  & 1.88 $\pm$ 0.04 & 11.6 $\pm$ 0.4 \\
HL-LHC & 36  & 2.12 $\pm$ 0.04 & 15.7 $\pm$ 0.4 \\
\hline
LHC & 300 & 5.41 $\pm$ 0.12 & 11.6 $\pm$ 0.4 \\
HL-LHC & 300 & 6.13 $\pm$ 0.11 & 15.7 $\pm$ 0.4 \\
\hline
LHC & 3000 & 17.12 $\pm$ 0.38 & 11.6 $\pm$ 0.4 \\
HL-LHC & 3000 & 19.39 $\pm$ 0.33 & 15.7 $\pm$ 0.4 \\
\hline
\hline
\end{tabular}
\label{lhc_sig}
\end{table}

The significance and $S/B$ decrease slightly in the LHC scenario, mainly because of the less performant $b$-tagging.
The results in the table indicate that $t\bar{t}H(H\rightarrow b\bar{b})$ could be observed already by the end of the LHC programme with an integrated luminosity of $300\ \rm{fb^{-1}}$ and a significance of $5.41 \pm 0.12$. 

\section{Top Yukawa Coupling Uncertainty}
\label{sec:yukawaerror}

The expected precision of a top Yukawa coupling ($y_t$) determination at the LHC and the HL-LHC was estimated from the uncertainty in the number of signal events. Values are shown in Table \ref{top_error}. The $t\bar{t}H$ cross section is proportional to the top Yukawa coupling squared, $\sigma_{t\bar{t}H} = k\, y_t^2$, where $k$ includes all the factors associated to a cross section computation. In this estimate $k$ and $\mathcal{L}$ are considered not to have associated errors.

\begin{table}[h!]
\centering
\caption{Relative uncertainty on the coupling of the Higgs boson to the top quark. Integrated luminosities of $3000\ \rm{fb^{-1}}$ and $3000\ \rm{fb^{-1}}$ are considered for the LHC and HL-LHC scenarios, respectively.}
\vspace{5pt}
\begin{tabular}{c|c|c}
\hline
\hline
Scenario & $\mathcal{L}$ ($\mbox{fb}^{-1}$) & $\Delta y_t/y_t$ ($\%$)\\ 
\hline
\hline
LHC & 300 & 35 \\
HL-LHC & 3000 & 17 \\
\hline
\hline
\end{tabular}
\label{top_error}
\end{table}


\section{Signal Strength}
\label{sec:signalmu}

The signal strength is obtained by minimizing $-2\,ln\,\lambda (\mu)$~\cite{pdg}, defined as 
\small{
\begin{equation}
\begin{split}
-2 \, ln \, \lambda (\mu) & = -2\, ln \frac{\mathcal{L(\mathbf{\mu}})}{\mathcal{L(\mathbf{\hat{\mu})}}} \\
& = 2 \sum_{i=1}^N \bigg[ (\mu\,s_i + b_i) - n_i + n_i\,ln \bigg( \frac{n_i}{\mu\,s_i + b_i} \bigg) \bigg] \nonumber
\end{split}
\end{equation}
}
\normalsize
where $N$ is the number of bins in the distribution, $\mathbf{\mu}$ is the signal strength and $\mathbf{\hat{\mu}}$ is the corresponding best estimator, $\mathcal{L(\mathbf{\mu})}$ is the likelihood estimator being maximized and $\mathcal{L(\mathbf{\hat{\mu})}}$ is the unconditional maximum likelihood estimator. Finally, $s_i$ and $b_i$ are the expected number of signal and background events, and $n_i$ is the number of events in the simulated $m_H$ distribution.

\begin{figure}[h!] 
\centering
\includegraphics[scale=0.35]{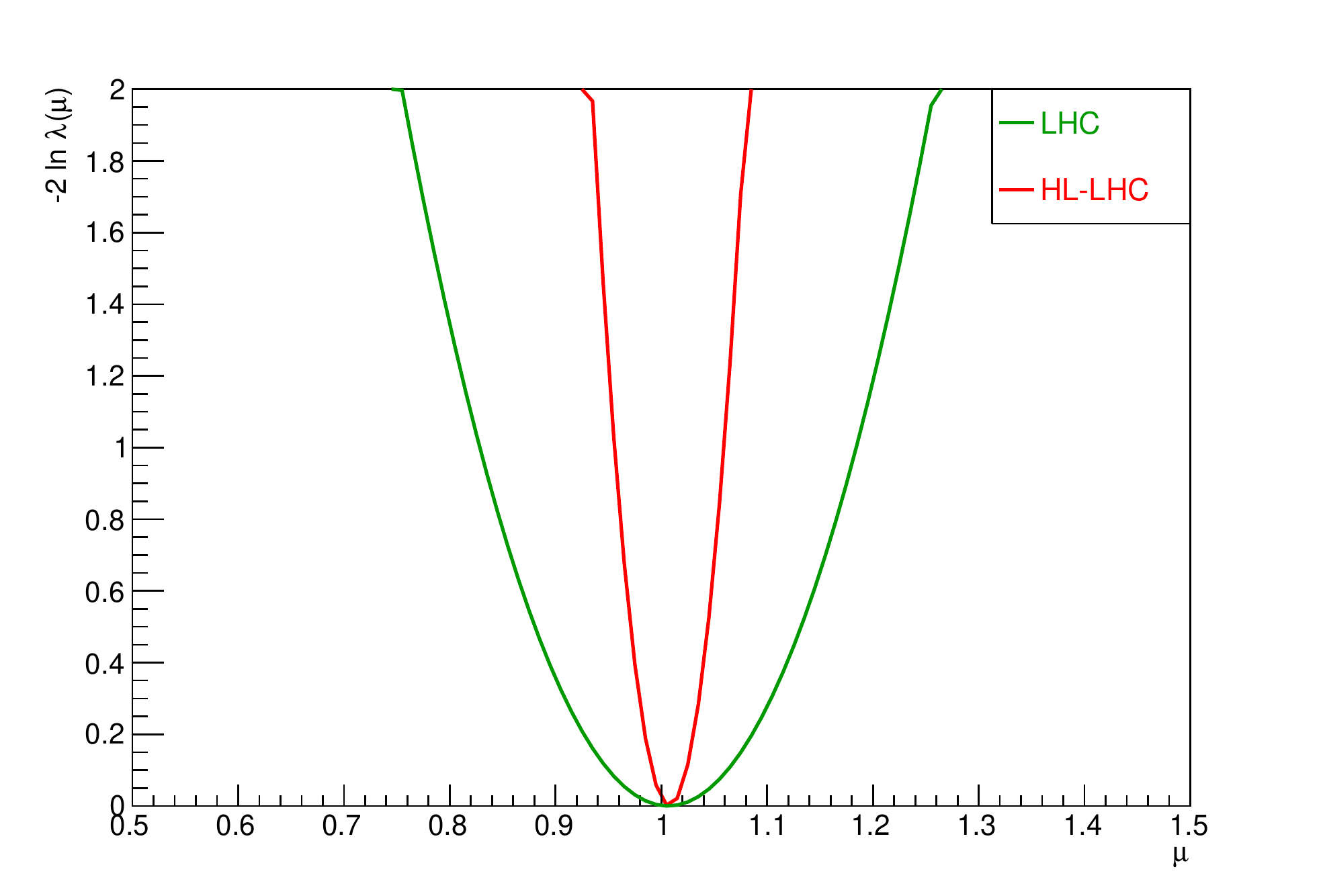}
\vspace{-9pt}
\caption{$-2\,ln\,\lambda (\mu)$ distribution for the LHC and HL-LHC scenarios, with an integrated luminosity of $300\ \rm{fb^{-1}}$ and $3\ \rm{ab^{-1}}$, respectively.}
\label{mu_fit}
\end{figure}

The distributions of $-2\,ln\,\lambda (\mu)$ for the LHC and HL-LHC scenarios, with an integrated luminosity of 300 fb$^{-1}$ and $3\ \rm{fb^{-1}}$, respectively, are presented in Figure~\ref{mu_fit}. As expected due to the larger integrated luminosity, it can be seen that the error on the signal strength decreases in the HL-LHC scenario.

The values for the signal strengths are shown in Table \ref{tab:mu_fit}. An uncertainty on the signal strength of 18\% is expected in the LHC scenario, while this error decreases to $5\%$ in the HL-LHC scenario. 

\begin{table}[h!]
\centering
\caption{Signal strength integrated for different luminosities and scenarios.}
\vspace{5pt}
\begin{tabular}{c|c|c}
\hline
\hline
Scenario & $\mathcal{L}$ ($\mbox{fb}^{-1}$) & Signal strength ($\mu$) \\ 
\hline
\hline
LHC & 300  & 0.99 $\pm$ 0.18 \\
HL-LHC & 3000  & 1.00 $\pm$ 0.05 \\
\hline
\hline
\end{tabular}
\label{tab:mu_fit}
\end{table}

\vspace*{-0.5cm}

\section{Search for a pure pseudo-scalar boson}

\begin{figure}[h!] 
\centering    
\includegraphics[scale=0.4]{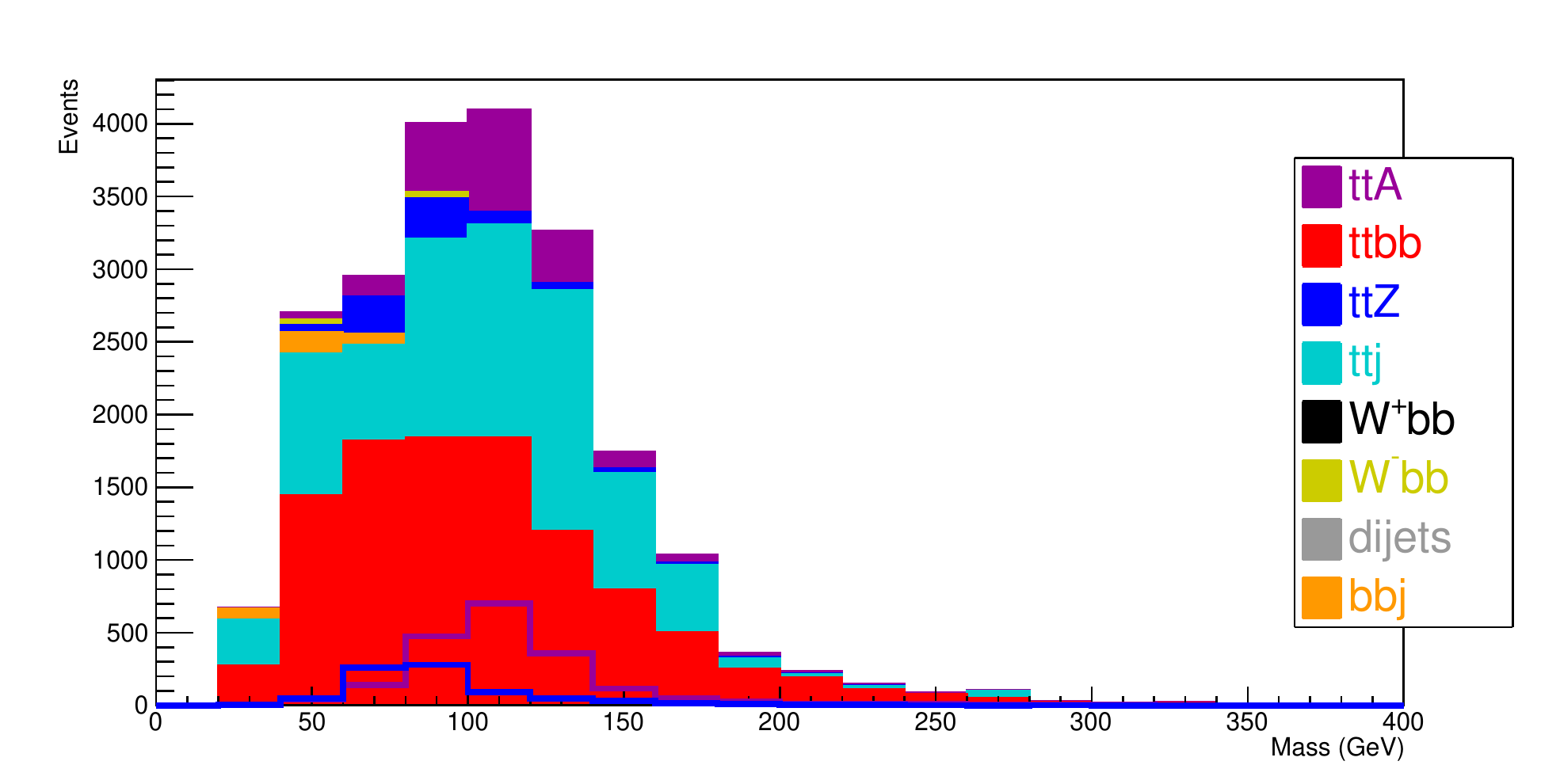}
\caption{Higgs candidates mass for $t\bar{t}A$ for $\mathcal{L} = 3 \,\mbox{ab}^{-1}$.}
\label{mass_ttA}
\end{figure}

The production of a $125\ \rm{GeV}$ pseudo-scalar $A$ in association with two top quarks was also considered in this work. Although it has been excluded that the observed $125\ \rm{GeV}$ Higgs is a pure pseudo-scalar, fermion vertices may yet uncover the presence of a pseudo-scalar component. The production cross section for $t\bar{t}A$ is about a half of the $t\bar{t}H$ one, and $A\rightarrow b\bar{b}$ decays were considered. The mass distribution of the Higgs candidate jets for the $t\bar{t}A$ signal sample and SM backgrounds, in the HL-LHC scenario, is shown in Figure~\ref{mass_ttA}.
The $t\bar{t}A$ and $t\bar{t}H$ distributions have similar shapes, differing only in the number of events due to the different cross sections.

The significance and $S/B$ for $t\bar{t}A$ production were computed for different integrated luminosities, and are shown in Table \ref{ttA_sig}. These variables are computed, as before, in the mass window between 60 and 160 GeV.

\begin{table}[h!]
\centering
\caption{Significance and S/B ratio for different integrated luminosities and processes computed from masses in range [60,160] GeV.}
\vspace{5pt}
\begin{tabular}{c|c|c|c}
\hline
\hline
Strategy & $\mathcal{L}$ ($\mbox{fb}^{-1}$) & $S/\sqrt{B}$ & $S/B$ ($\%$)\\ 
\hline
\hline
$t\bar{t}H$ & 36  & 2.12 $\pm$ 0.04 & 15.7 $\pm$ 0.4 \\
$t\bar{t}A$ & 36  & 1.63 $\pm$ 0.03 & 12.1 $\pm$ 0.3 \\
\hline
$t\bar{t}H$ & 300 & 6.13 $\pm$ 0.11 & 15.7 $\pm$ 0.4 \\
$t\bar{t}A$ & 300 & 4.71 $\pm$ 0.10 & 12.1 $\pm$ 0.3 \\
\hline
$t\bar{t}H$ & 3000 & 19.39 $\pm$ 0.33 & 15.7 $\pm$ 0.4 \\
$t\bar{t}A$ & 3000 & 14.90 $\pm$ 0.32 & 12.1 $\pm$ 0.3 \\
\hline
\hline
\end{tabular}
\label{ttA_sig}
\end{table}

The estimated significance and $S/B$ are lower for $t\bar{t}A$ production. Observation of $t\bar{t}A$ production would require at least $\mathcal{L}=350 fb^{-1}$ in the HL-LHC scenario, with an expected significance of $5.09 \pm 0.10$.

\section{Conclusions}
\label{sec:concl}

An analysis strategy for the semileptonic $t\bar{t}H (H\rightarrow b\bar{b})$ channel is proposed in this paper. It relies on the reconstruction of boosted Higgs bosons using large radius jets and jet substructure information to identify the objects of interest and suppress backgrounds. 

It improves the analysis significance by a factor 3 with respect to Reference~\cite{mangano} (which was optimized for the Future Circular Collider). 
Moreover, it was observed that the re-clustering technique may be used without affecting the results. Finally, a control region is proposed, kinematically close to the signal region, although orthogonal to it through the use of anti-$b$-tagging.

Results indicate that 
$t\bar{t}H(H\rightarrow b\bar{b})$ could be observed at the LHC with an integrated luminosity of 300 fb$^{-1}$ in the LHC scenario, using the optimized strategy, with a significance of $5.41\pm 0.12$.

The top Yukawa coupling extracted from the proposed analysis is expected to have a 35\% uncertainty by the end of the LHC programme, considering an integrated luminosity of $300\ \rm{fb^{-1}}$. This uncertainty decreases to 17$\%$ in the HL-LHC scenario with an integrated luminosity of $3\ \rm{ab^{-1}}$.

A multivariate method (MVA) could further discriminate between the signal and the backgrounds, exploiting correlations between discriminating variables such as $\tau_{21}$ and $\tau_{31}$ ratios for the Higgs candidate jets.
Finally, it should be noted that this paper does not consider the effects of pile-up, neither uses a full simulation of the detector.
The analysis sensitivity is expected to decrease when introducing these realistic effects, but continue to be competitive in terms of significance. 

\vspace*{0.5cm}
\section*{Acknowledgements}

The authors would like to thank our colleagues and friends Liliana Apolinário, Pedro Abreu, Jo\~{a}o Martins, Aidan Kelly, and Silvia Biondi, for various discussions, support, and good advice, as well as for part of the $b\bar{b}j$ and di-jet samples used. This work was partially supported by Funda\c{c}\~{a}o para a Ci\^{e}ncia e Tecnologia, FCT (Projects No. CERN/FIS-PAR/0008/2017).


\bibliography{ttHbb_note}

\begin{thebibliography}{23}%
\makeatletter
\providecommand \@ifxundefined [1]{%
 \@ifx{#1\undefined}
}%
\providecommand \@ifnum [1]{%
 \ifnum #1\expandafter \@firstoftwo
 \else \expandafter \@secondoftwo
 \fi
}%
\providecommand \@ifx [1]{%
 \ifx #1\expandafter \@firstoftwo
 \else \expandafter \@secondoftwo
 \fi
}%
\providecommand \natexlab [1]{#1}%
\providecommand \enquote  [1]{``#1''}%
\providecommand \bibnamefont  [1]{#1}%
\providecommand \bibfnamefont [1]{#1}%
\providecommand \citenamefont [1]{#1}%
\providecommand \href@noop [0]{\@secondoftwo}%
\providecommand \href [0]{\begingroup \@sanitize@url \@href}%
\providecommand \@href[1]{\@@startlink{#1}\@@href}%
\providecommand \@@href[1]{\endgroup#1\@@endlink}%
\providecommand \@sanitize@url [0]{\catcode `\\12\catcode `\$12\catcode
  `\&12\catcode `\#12\catcode `\^12\catcode `\_12\catcode `\%12\relax}%
\providecommand \@@startlink[1]{}%
\providecommand \@@endlink[0]{}%
\providecommand \url  [0]{\begingroup\@sanitize@url \@url }%
\providecommand \@url [1]{\endgroup\@href {#1}{\urlprefix }}%
\providecommand \urlprefix  [0]{URL }%
\providecommand \Eprint [0]{\href }%
\providecommand \doibase [0]{http://dx.doi.org/}%
\providecommand \selectlanguage [0]{\@gobble}%
\providecommand \bibinfo  [0]{\@secondoftwo}%
\providecommand \bibfield  [0]{\@secondoftwo}%
\providecommand \translation [1]{[#1]}%
\providecommand \BibitemOpen [0]{}%
\providecommand \bibitemStop [0]{}%
\providecommand \bibitemNoStop [0]{.\EOS\space}%
\providecommand \EOS [0]{\spacefactor3000\relax}%
\providecommand \BibitemShut  [1]{\csname bibitem#1\endcsname}%
\let\auto@bib@innerbib\@empty
\bibitem [{\citenamefont {Collaboration}(2012{\natexlab{a}})}]{higgsAtlas}%
  \BibitemOpen
  \bibfield  {author} {\bibinfo {author} {\bibfnamefont {A.}~\bibnamefont
  {Collaboration}},\ }\href@noop {} {\bibfield  {journal} {\bibinfo  {journal}
  {Physics Letters B}\ }\textbf {\bibinfo {volume} {716}},\ \bibinfo {pages}
  {1–29} (\bibinfo {year} {2012}{\natexlab{a}})},\ \bibinfo {note}
  {arXiv$:$1207.7214 [hep-ex]}\BibitemShut {NoStop}%
\bibitem [{\citenamefont {Collaboration}(2012{\natexlab{b}})}]{higgsCMS}%
  \BibitemOpen
  \bibfield  {author} {\bibinfo {author} {\bibfnamefont {C.}~\bibnamefont
  {Collaboration}},\ }\href@noop {} {\bibfield  {journal} {\bibinfo  {journal}
  {Physics Letters B}\ }\textbf {\bibinfo {volume} {716}},\ \bibinfo {pages}
  {30} (\bibinfo {year} {2012}{\natexlab{b}})},\ \bibinfo {note}
  {arXiv$:$1207.7235 [hep-ex]}\BibitemShut {NoStop}%
\bibitem [{\citenamefont {$\textit{et al.}$
  (ATLAS~Collaboration)}(2018{\natexlab{a}})}]{ttHobs}%
  \BibitemOpen
  \bibfield  {author} {\bibinfo {author} {\bibfnamefont {M.~A.}\ \bibnamefont
  {$\textit{et al.}$ (ATLAS~Collaboration)}},\ }\href@noop {} {\bibfield
  {journal} {\bibinfo  {journal} {Phys. Lett. B}\ }\textbf {\bibinfo {volume}
  {784}},\ \bibinfo {pages} {173 } (\bibinfo {year} {2018}{\natexlab{a}})},\
  \bibinfo {note} {arXiv:1806.00425v1 [hep-ex]}\BibitemShut {NoStop}%
\bibitem [{\citenamefont {$\textit{et al.}$
  (CMS~Collaboration)}(2018{\natexlab{a}})}]{ttHobs2}%
  \BibitemOpen
  \bibfield  {author} {\bibinfo {author} {\bibfnamefont {A.~S.}\ \bibnamefont
  {$\textit{et al.}$ (CMS~Collaboration)}},\ }\href@noop {} {\bibfield
  {journal} {\bibinfo  {journal} {Phys. Rev. Lett.}\ }\textbf {\bibinfo
  {volume} {120}},\ \bibinfo {pages} {231801} (\bibinfo {year}
  {2018}{\natexlab{a}})},\ \bibinfo {note} {arXiv:1804.02610v2
  [hep-ex]}\BibitemShut {NoStop}%
\bibitem [{\citenamefont {$\textit{et al.}$
  (ATLAS~Collaboration)}(2018{\natexlab{b}})}]{bbobs}%
  \BibitemOpen
  \bibfield  {author} {\bibinfo {author} {\bibfnamefont {M.~A.}\ \bibnamefont
  {$\textit{et al.}$ (ATLAS~Collaboration)}},\ }\href@noop {} {\  (\bibinfo
  {year} {2018}{\natexlab{b}})},\ \bibinfo {note} {arXiv:1808.08238v1 [hep-ex],
  CERN-EP-2018-215}\BibitemShut {NoStop}%
\bibitem [{\citenamefont {$\textit{et al.}$
  (CMS~Collaboration)}(2018{\natexlab{b}})}]{bbobs2}%
  \BibitemOpen
  \bibfield  {author} {\bibinfo {author} {\bibfnamefont {A.~S.}\ \bibnamefont
  {$\textit{et al.}$ (CMS~Collaboration)}},\ }\href@noop {} {\  (\bibinfo
  {year} {2018}{\natexlab{b}})},\ \bibinfo {note} {arXiv:1808.08242v1 [hep-ex],
  CMS-PAS-HIG-18-016, CERN-EP-2018-223}\BibitemShut {NoStop}%
\bibitem [{\citenamefont {$\textit{et al.}$
  (ATLAS~Collaboration)}(2018{\natexlab{c}})}]{tauobs}%
  \BibitemOpen
  \bibfield  {author} {\bibinfo {author} {\bibfnamefont {M.~A.}\ \bibnamefont
  {$\textit{et al.}$ (ATLAS~Collaboration)}},\ }\href@noop {} {\  (\bibinfo
  {year} {2018}{\natexlab{c}})},\ \bibinfo {note}
  {aTLAS-CONF-2018-021}\BibitemShut {NoStop}%
\bibitem [{\citenamefont {$\textit{et al.}$
  (CMS~Collaboration)}(2018{\natexlab{c}})}]{tauobs2}%
  \BibitemOpen
  \bibfield  {author} {\bibinfo {author} {\bibfnamefont {A.~S.}\ \bibnamefont
  {$\textit{et al.}$ (CMS~Collaboration)}},\ }\href@noop {} {\bibfield
  {journal} {\bibinfo  {journal} {Phys. Lett. B}\ }\textbf {\bibinfo {volume}
  {779}},\ \bibinfo {pages} {283} (\bibinfo {year} {2018}{\natexlab{c}})},\
  \bibinfo {note} {arXiv:1708.00373v2 [hep-ex]}\BibitemShut {NoStop}%
\bibitem [{\citenamefont {Grojean}(2017)}]{brcs}%
  \BibitemOpen
  \bibfield  {author} {\bibinfo {author} {\bibfnamefont {C.}~\bibnamefont
  {Grojean}},\ }\href@noop {} {\bibfield  {journal} {\bibinfo  {journal} {CERN
  Yellow Report}\ ,\ \bibinfo {pages} {143}} (\bibinfo {year} {2017})},\
  \bibinfo {note} {cERN 2016-005, arXiv$:$1708.00794 [hep-ph]}\BibitemShut
  {NoStop}%
\bibitem [{\citenamefont {Collaboration}(2018)}]{ttH}%
  \BibitemOpen
  \bibfield  {author} {\bibinfo {author} {\bibfnamefont {A.}~\bibnamefont
  {Collaboration}},\ }\href@noop {} {\bibfield  {journal} {\bibinfo  {journal}
  {Phys. Rev. D}\ }\textbf {\bibinfo {volume} {97}},\ \bibinfo {pages} {072016}
  (\bibinfo {year} {2018})},\ \bibinfo {note} {cERN-EP-2017-291,
  arXiv$:$1712.08895 [hep-ex]}\BibitemShut {NoStop}%
\bibitem [{\citenamefont {M.~L.~Mangano}\ \emph {et~al.}(2016)\citenamefont
  {M.~L.~Mangano}, \citenamefont {Schell},\ and\ \citenamefont
  {Shao}}]{mangano}%
  \BibitemOpen
  \bibfield  {author} {\bibinfo {author} {\bibfnamefont {P.~R.}\ \bibnamefont
  {M.~L.~Mangano}, \bibfnamefont {T.~Plehn}}, \bibinfo {author} {\bibfnamefont
  {T.}~\bibnamefont {Schell}}, \ and\ \bibinfo {author} {\bibfnamefont
  {H.}~\bibnamefont {Shao}},\ }\href@noop {} {\bibfield  {journal} {\bibinfo
  {journal} {Journal of Physics G: Nuclear and Particle Physics}\ }\textbf
  {\bibinfo {volume} {43}},\ \bibinfo {pages} {035001} (\bibinfo {year}
  {2016})},\ \bibinfo {note} {arXiv$:$1507.08169v2 [hep-ph]}\BibitemShut
  {NoStop}%
\bibitem [{\citenamefont {Collaboration}(2015)}]{HL}%
  \BibitemOpen
  \bibfield  {author} {\bibinfo {author} {\bibfnamefont {A.}~\bibnamefont
  {Collaboration}},\ }\href@noop {} {\  (\bibinfo {year} {2015})},\ \bibinfo
  {note} {cERN-LHCC-2015-020, LHCC-G-166}\BibitemShut {NoStop}%
\bibitem [{\citenamefont {G.~Apollinari}\ \emph {et~al.}(2017)\citenamefont
  {G.~Apollinari}, \citenamefont {Fessia}, \citenamefont {Lamont},
  \citenamefont {Rossi},\ and\ \citenamefont {Tavian}}]{HL3}%
  \BibitemOpen
  \bibfield  {author} {\bibinfo {author} {\bibfnamefont {O.~B.~O.}\
  \bibnamefont {G.~Apollinari}, \bibfnamefont {I.~Béjar~Alonso}}, \bibinfo
  {author} {\bibfnamefont {P.}~\bibnamefont {Fessia}}, \bibinfo {author}
  {\bibfnamefont {M.}~\bibnamefont {Lamont}}, \bibinfo {author} {\bibfnamefont
  {L.}~\bibnamefont {Rossi}}, \ and\ \bibinfo {author} {\bibfnamefont
  {L.}~\bibnamefont {Tavian}},\ }\href@noop {} {\  (\bibinfo {year} {2017})},\
  \bibinfo {note} {cERN-2017-007-M}\BibitemShut {NoStop}%
\bibitem [{\citenamefont {$\textit{et al.}$}(2013)}]{ttamodel}%
  \BibitemOpen
  \bibfield  {author} {\bibinfo {author} {\bibfnamefont {P.~A.}\ \bibnamefont
  {$\textit{et al.}$}},\ }\href@noop {} {\bibfield  {journal} {\bibinfo
  {journal} {JHEP}\ }\textbf {\bibinfo {volume} {2013}},\ \bibinfo {pages} {43}
  (\bibinfo {year} {2013})},\ \bibinfo {note} {arXiv:1306.6464v3
  [hep-ph]}\BibitemShut {NoStop}%
\bibitem [{\citenamefont {$\textit{et al.}$}(2014)}]{mg5}%
  \BibitemOpen
  \bibfield  {author} {\bibinfo {author} {\bibfnamefont {J.~A.}\ \bibnamefont
  {$\textit{et al.}$}},\ }\href@noop {} {\bibfield  {journal} {\bibinfo
  {journal} {JHEP}\ }\textbf {\bibinfo {volume} {2014}},\ \bibinfo {pages} {79}
  (\bibinfo {year} {2014})},\ \bibinfo {note} {arXiv$:$ 1405.0301
  [hep-ph]}\BibitemShut {NoStop}%
\bibitem [{\citenamefont {T.~Sjostrand}\ and\ \citenamefont
  {Skands}(2008)}]{pythia}%
  \BibitemOpen
  \bibfield  {author} {\bibinfo {author} {\bibfnamefont {S.~M.}\ \bibnamefont
  {T.~Sjostrand}}\ and\ \bibinfo {author} {\bibfnamefont {P.~Z.}\ \bibnamefont
  {Skands}},\ }\href@noop {} {\bibfield  {journal} {\bibinfo  {journal}
  {Computer Physics Communications}\ }\textbf {\bibinfo {volume} {178}},\
  \bibinfo {pages} {852} (\bibinfo {year} {2008})},\ \bibinfo {note}
  {arXiv$:$0710.3820 [hep-ph]}\BibitemShut {NoStop}%
\bibitem [{\citenamefont {P.~Artoisenet}\ and\ \citenamefont
  {Rietkerk}(2013)}]{madspin}%
  \BibitemOpen
  \bibfield  {author} {\bibinfo {author} {\bibfnamefont {O.~M.}\ \bibnamefont
  {P.~Artoisenet}, \bibfnamefont {R.~Frederix}}\ and\ \bibinfo {author}
  {\bibfnamefont {R.}~\bibnamefont {Rietkerk}},\ }\href@noop {} {\bibfield
  {journal} {\bibinfo  {journal} {JHEP}\ }\textbf {\bibinfo {volume} {2013}},\
  \bibinfo {pages} {15} (\bibinfo {year} {2013})},\ \bibinfo {note}
  {arXiv$:$1212.3460 [hep-ph]}\BibitemShut {NoStop}%
\bibitem [{\citenamefont {de~Favereau~$\textit{et al.}$ (DELPHES
  3~Collaboration)}(2014)}]{delphes}%
  \BibitemOpen
  \bibfield  {author} {\bibinfo {author} {\bibfnamefont {J.}~\bibnamefont
  {de~Favereau~$\textit{et al.}$ (DELPHES 3~Collaboration)}},\ }\href@noop {}
  {\bibfield  {journal} {\bibinfo  {journal} {JHEP}\ }\textbf {\bibinfo
  {volume} {2014}},\ \bibinfo {pages} {57} (\bibinfo {year} {2014})},\ \bibinfo
  {note} {arXiv$:$1307.6346 [hep-ex]}\BibitemShut {NoStop}%
\bibitem [{\citenamefont {$\textit{et al.}$
  (ATLAS~Collaboration)}(2017)}]{pixelbtag}%
  \BibitemOpen
  \bibfield  {author} {\bibinfo {author} {\bibfnamefont {M.~A.}\ \bibnamefont
  {$\textit{et al.}$ (ATLAS~Collaboration)}},\ }\href@noop {} {\  (\bibinfo
  {year} {2017})},\ \bibinfo {note} {aTLAS-TDR-030,
  CERN-LHCC-2017-021}\BibitemShut {NoStop}%
\bibitem [{\citenamefont {Y.~L.~Dokshitzer}\ and\ \citenamefont
  {Webber}(1997)}]{ca}%
  \BibitemOpen
  \bibfield  {author} {\bibinfo {author} {\bibfnamefont {S.~M.}\ \bibnamefont
  {Y.~L.~Dokshitzer}, \bibfnamefont {G.~D.~Leder}}\ and\ \bibinfo {author}
  {\bibfnamefont {B.~R.}\ \bibnamefont {Webber}},\ }\href@noop {} {\bibfield
  {journal} {\bibinfo  {journal} {JHEP}\ }\textbf {\bibinfo {volume} {1997}},\
  \bibinfo {pages} {001} (\bibinfo {year} {1997})},\ \bibinfo {note}
  {arXiv:hep-ph/9707323}\BibitemShut {NoStop}%
\bibitem [{\citenamefont {J.~M.~Butterworth}\ and\ \citenamefont
  {Salam}(2008)}]{bdrs}%
  \BibitemOpen
  \bibfield  {author} {\bibinfo {author} {\bibfnamefont {M.~R.}\ \bibnamefont
  {J.~M.~Butterworth}, \bibfnamefont {A.~R.~Davison}}\ and\ \bibinfo {author}
  {\bibfnamefont {G.~P.}\ \bibnamefont {Salam}},\ }\href@noop {} {\bibfield
  {journal} {\bibinfo  {journal} {Phys. Rev. Lett.}\ }\textbf {\bibinfo
  {volume} {100}},\ \bibinfo {pages} {242001} (\bibinfo {year} {2008})},\
  \bibinfo {note} {arXiv:0802.2470v2 [hep-ph]}\BibitemShut {NoStop}%
\bibitem [{\citenamefont {T.~Plehn}\ and\ \citenamefont
  {Spannowsky}(2010)}]{toptag}%
  \BibitemOpen
  \bibfield  {author} {\bibinfo {author} {\bibfnamefont {G.~P.~S.}\
  \bibnamefont {T.~Plehn}}\ and\ \bibinfo {author} {\bibfnamefont
  {M.}~\bibnamefont {Spannowsky}},\ }\href@noop {} {\bibfield  {journal}
  {\bibinfo  {journal} {Phys. Rev. Lett.}\ }\textbf {\bibinfo {volume} {104}},\
  \bibinfo {pages} {111801} (\bibinfo {year} {2010})},\ \bibinfo {note}
  {arXiv:0910.5472v2 [hep-ph]}\BibitemShut {NoStop}%
\bibitem [{\citenamefont {et~al. (Particle Data~Group)}(2016)}]{pdg}%
  \BibitemOpen
  \bibfield  {author} {\bibinfo {author} {\bibfnamefont {C.~P.}\ \bibnamefont
  {et~al. (Particle Data~Group)}},\ }\href@noop {} {\enquote {\bibinfo {title}
  {{Particle Physics Booklet}},}\ } (\bibinfo {year} {2016}),\ \bibinfo {note}
  {chin. Phys. C, 40, 100001}\BibitemShut {NoStop}%
\end{thebibliography}%
\end{document}